\documentstyle[preprint,aps,prd,eqsecnum]{revtex}
\newcommand{\gsim}{
\raisebox{-4pt}{$\,\stackrel{\textstyle >}{\sim}\,$}}
\begin{document}
\preprint{WU~B~94-28}
\draft
\title{MESON-PHOTON TRANSITION FORM FACTORS}
\author{R.~Jakob, P.~Kroll\footnote
{kroll@wpts0.physik.uni-wuppertal.de}, M.~Raulfs}
\address{Fachbereich Physik, Universit\"at Wuppertal,\\
D-42097 Wuppertal, Germany}
\date{October 17, 1994}
\maketitle
\begin{abstract}
We report results on the $\pi$-$\gamma$ transition form
factor obtained within the hard scattering
approach including transverse momentum effects and
Sudakov corrections. The results clearly favor distribution
amplitudes close to the asymptotic form, $\sim x_1x_2$, and
disfavor distribution amplitudes which are strongly
concentrated in the end-point regions. This observation is backed by
information on the elastic form factor of the pion and on its valence quark
distribution function. Applications of our approach to the $\eta$-$\gamma$ and
$\eta^\prime$-$\gamma$ transition form factors are discussed as
well. Combining the form factor data with the two-photon decay widths, we
determine the $\eta$ and the $\eta^\prime$ decay constants and
the $\eta$-$\eta^\prime$ mixing angle.
\end{abstract}
\newpage

%%%%%%%%%%%%%%%%%%%%%%%%%%%%%%%%%%%%%%%%%%%%%%%%%%%%%%%%%%%%%%%%%%%%%%%%%%%

\section{introduction}
\label{sec:intro}

Hadronic form factors at large momentum transfer provide information on the
constituents the hadrons are made of and on the dynamics controlling their
interactions. Therefore, the form factors always found much interest and
many papers, both theoretical and experimental ones, are devoted to
them. Recently a new perturbative approach has been proposed by Botts,
Li and Sterman \cite{Bot:89,LiS:92,Li:92} which allows to calculate
the large momentum behavior of form factors.
In this new approach, which one may term the modified hard scattering
approach (HSA), the transverse degrees of freedom as well as gluonic radiative
corrections---condensed in a Sudakov factor---are taken into account
in contrast to the standard perturbative approach \cite{Lep:80}.
Applications of the modified HSA to the pion's and nucleon's form
factor\cite{LiS:92,Li:92,Jak:93,Bol:94} revealed that the perturbative
contributions to these form factors can reliably and
self-consistently (in the sense that the bulk of the contributions is
accumulated in regions where the strong coupling $\alpha_s$ is
sufficiently small) be calculated. It turned out, however, that
the perturbative contributions are too small as compared with the
data. Responsible for that discrepancy might be omitted
contributions
from higher order perturbation theory and/or from higher Fock states.
Another and perhaps the most important source of additional
contributions to the form factors are
the overlap of the soft wave functions
for the initial and final state hadrons \cite{Dre:70}. All these
contributions are inherent to the standard as well as to the modified
HSA but, with very few exceptions, have not been
considered as yet.\\
In this paper we are going to apply the modified HSA to pseudoscalar
meson-photon transition form factors for which data in the few GeV region
is now available \cite{Ber:84,Aih:90,Beh:91}. These transition
form factors are
exceptional cases since there is no overlap contribution in contrast to
the elastic hadronic form factors and, to lowest order, they are QED
processes. QCD should only provide corrections of the order of
$10-20\%$ and higher Fock state contributions are suppressed
by additional powers of $\alpha_s/Q^2$. For these reasons
one may expect the modified HSA to work for momentum transfer $Q$
larger than about $1\,{\rm GeV}$. The low $Q$ behavior of the
transition form factors has been investigated in a Salpeter model
in \cite{Mue:94}.\\
Input to calculations within the modified HSA are the hadronic
wave functions which contain the long-distance physics and are not
calculable at present. However, the pion wave function required
for the calculation of the $\pi$-$\gamma$ transition form factor,
$F_{\pi\gamma}(Q^2)$, is rather well constrained. A reliable
quantitative estimate of
the $\pi$-$\gamma$ transition form factor can therefore be made, as
we will discuss in Sec.~\ref{sec:pigaff}. We extend this calculation in
Sec.~\ref{sec:etagaff} to the $\eta$-$\gamma$ and $\eta^\prime$-$\gamma$
form factors and determine the decay constants and the mixing angle for
pseudoscalar mesons. Sec.~\ref{sec:piff} is
devoted to a discussion of the elastic form factor of the pion and to its
valence quark distribution function. It will turn out that the same pion
wave function as is used in the calculation of the $\pi$-$\gamma$
transition form factor, also leads to a reasonable description of
the valence quark distribution function and the elastic pion form
factor provided the overlap contribution is also taken into account in
the latter case. We recall that the Drell-Yan-West overlap formula
\cite{Dre:70} is the starting point of the derivation of the hard scattering
formula for the elastic pion form factor. The paper terminates
with a few concluding remarks (Sec.~\ref{sec:concl}).

%%%%%%%%%%%%%%%%%%%%%%%%%%%%%%%%%%%%%%%%%%%%%%%%%%%%%%%%%%%%%%%%%%%%%%%%

\section{the $\pi$-$\gamma$ transition form factor}
\label{sec:pigaff}

Adapting the modified HSA to the case of $\pi$-$\gamma$ transitions
we write the corresponding form factor as
\begin{equation}
F_{\pi\gamma}(Q^2)=
\int dx_1 \frac{d^{\;\!2}b_1}{4\pi} \;
\hat \Psi_0(x_1,-\vec b_1) \;
\hat T_H(x_1,\vec b_1,Q) \;
\exp\left[ -S(x_1,b_1,Q) \right],
\label{eq:pigaff-ft}
\end{equation}
where $\vec b_1$ is the quark-antiquark separation in the transverse
configuration space. $x_1$ and $x_2=1-x_1$ denote the usual momentum
fractions the quark and the antiquark carry, respectively. $\hat T_H$ is
the Fourier transform of the momentum space hard scattering amplitude
to be calculated from the Feynman diagrams shown in Fig.~\ref{fig:diags}.
Neglecting the quark masses and the mass of the pion, the hard scattering
amplitude $T_H$ in momentum space reads
\begin{equation}
T_H(x_1,\vec k_\perp,Q) =
2\sqrt{6} \, C_\pi \;
\left\{
\frac{1}{x_2Q^2 + k_\perp^2}
+\frac{1}{x_1Q^2 + k_\perp^2}
\right\},
\label{eq:pigaff-TH}
\end{equation}
where the charge factor $C_\pi$ is $({e_u}^2-{e_d}^2)/\sqrt{2}$. $e_i$ denotes
the charge of quark $i$ in units of the elementary charge.
The Fourier transform of this amplitude reads
\begin{equation}
\hat T_H(x_1,\vec b_1,Q) =
\frac{2\sqrt{6}\,C_\pi}{\pi}
K_0(\sqrt{x_2}Qb_1),
\label{eq:pigaff-TH-ft}
\end{equation}
where $K_0$ is the modified Bessel function of order zero.
Strictly speaking, (\ref{eq:pigaff-TH-ft}) represents twice
the Fourier transform of the first term in (\ref{eq:pigaff-TH}).
Because of the symmetry of the pion wave function under the
replacements $x_1\leftrightarrow x_2$, which appears as a consequence
of charge
conjugation invariance, the second term leads to the same contribution as
the first term after integration over the transverse separation
$\vec b_1$. The Sudakov exponent $S$ in (\ref{eq:pigaff-ft}) comprising the
gluonic radiative corrections, is given by
\begin{equation}
S(x_1,Q,b_1)= s(x_1,Q,b_1)+s(x_2,Q,b_1)
-\frac{4}{\beta_0} \ln \frac{\ln (t/\Lambda_{QCD})}
{\ln (1/b_1 \Lambda_{QCD})}
\label{eq:pigaff-sudakov}
\end{equation}
where a Sudakov function $s$ appears for each quark
line entering the hard scattering amplitude. The last term in
(\ref{eq:pigaff-sudakov}) arises from the application of the renormalization
group equation ($\beta_0=11-2/3 \,n_f$). A value of $200\,{\rm MeV}$ for
$\Lambda_{QCD}$ is used throughout and $t$ is taken to be the
largest mass scale appearing in $T_H$,
i.~e., $t=\max(\sqrt{x_2}Q,1/b_1)$. For small $b_1$ there is no
suppression from the Sudakov factor; as $b_1$ increases the Sudakov
factor decreases, reaching zero at $b_1=1/\Lambda_{QCD}$. For even
larger $b_1$ the Sudakov factor is set to zero. The Sudakov function $s$
is explicitly given in \cite{Bot:89,LiS:92}.\\
The quantity $\hat \Psi_0$ appearing in (\ref{eq:pigaff-ft}) represents
the soft part of the
transverse configuration space pion wave function, i.~e.,
the full wave function with the perturbative tail removed from
it. Following \cite{Jak:93} we write the wave function as
\begin{equation}
\hat\Psi_0 (x_1,\vec b_1) = \frac{f_\pi}{2 \sqrt{6}}
\,\phi(x_1) \,\hat\Sigma(\sqrt{x_1x_2}\,b_1).
\label{eq:wvfct-ansatz}
\end{equation}
It is subject to the auxiliary conditions
\begin{equation}
\hat\Sigma ( 0 ) =4\pi, \qquad\qquad
\int_0^1 dx_1 \;\phi(x_1) =1.
\label{eq:auxiliary}
\end{equation}
The wave function does not factorize in $x_1$ and $b_1$, but
in accord with the basic properties of the HSA \cite{Bro:83,Zhi:93}
the $b_1$-dependence rather appears in the combination
$\sqrt{x_1x_2}\,b_1$. The transverse part of the wave function
is assumed to be a simple Gaussian
\begin{equation}
\hat\Sigma(\sqrt{x_1x_2}\,b_1)=
4\pi \;
\exp \left( -x_1x_2\,b_1^2/4 a^2 \right).
\label{eq:Sigma-ft}
\end{equation}
In \cite{Bro:83} it is shown how (\ref{eq:Sigma-ft}) is
related to the equal time harmonic oscillator
wave  function. More complicated forms than (\ref{eq:Sigma-ft})
(e.~g., a two-humped shape of the momentum space wave function)
are proposed in \cite{Zhi:93} on the basis of dispersion relations
and duality. At large transverse
momentum, however, the soft momentum space
wave function should behave like a Gaussian \cite{Zhi:93}. The
examination of a number of examples corroborates our
expectation that forms of $\hat\Sigma$ other than (\ref{eq:Sigma-ft})
will not change the results and the conclusions presented in our
paper markedly.\\
The two free parameters contained in our ansatz, namely the value
of the wave function at the origin and the parameter $a$ controlling
the transverse size, are well
fixed by the decay processes $\pi^+\to\mu^+\nu_\mu$ and
$\pi^0\to\gamma\gamma$ providing the relations \cite{Bro:83}
\begin{equation}
\int_0^1 dx_1 \; \hat\Psi_0(x_1,\vec b_1=0) =\frac{2\pi\,f_\pi}{\sqrt{6}},
\qquad \qquad
\int dx_1\,d^{\;\!2}b_1 \; \hat\Psi_0(x_1,\vec b_1)=\frac{\sqrt{6}}{f_\pi},
\label{eq:pimunupigamgam}
\end{equation}
where $f_\pi(=130.7\,{\rm MeV})$ is the usual pion decay constant.
The first relation is automatically satisfied by our ansatz
(\ref{eq:wvfct-ansatz}), whereas the second relation is used to fix the
parameter $a$. For the distribution
amplitude, $\phi$, we use the asymptotic form
\begin{equation}
\label{eq:DA-AS}
\phi_{AS}(x_1)=6\, x_1x_2
\end{equation}
and alternatively, as a representative of strongly end-point concentrated
distribution amplitudes, a form proposed by Chernyak and
Zhitnitsky \cite{Che:82}
\begin{equation}
\label{eq:DA-CZ}
\phi_{CZ}(x_1)=30\, x_1x_2 \, (x_1-x_2)^2 .
\end{equation}
It can be shown (see \cite{Lep:80} and references therein) that
the distribution amplitudes are subject to evolution and
can be expanded over Gegenbauer polynomials which are the eigenfunctions
of the evolution equation for mesons
\begin{equation}
\phi(x_1)=\phi_{AS}(x_1) \,
\left[
1+\sum_{n=2}^\infty B_n \,
\left(\frac{\alpha_s(\mu_F)}{\alpha_s(\mu_0)}\right)^{\gamma_n} \,
\, C_n^{3/2}(x_1 -x_2)
\right] .
\label{eq:evolution}
\end{equation}
$\mu_F$ is a scale of order $Q$ at which soft
and hard physics factorizes and $\mu_0$ is a typical hadronic scale of order
$1\,{\rm GeV}$. Charge conjugation invariance requires the odd $n$ expansion
coefficients $B_n$ to vanish. Since the $\gamma_n$ are positive
fractional numbers any distribution amplitude evolves into
$\phi_{AS}(x)$ asymptotically, i.~e., for $\ln (Q/\mu_0) \to \infty$. The
asymptotic distribution amplitude itself shows no evolution. In the
representation (\ref{eq:evolution}) the CZ distribution amplitude
is given by $B_2=2/3$ and $B_n=0$ for $n>2$. Its evolution
can safely be ignored in the very limited range of momentum transfer
we are interested in.\\
In the standard HSA the distribution amplitude (\ref{eq:DA-CZ}) leads
to a prediction for the elastic pion form
factor in apparent agreement with experiment at the expense, however,
of the dominance of contributions from the end-point
regions, $x_1\to 0$ or $1$, where the use of the perturbative QCD
is unjustified
as has been pointed out by several authors \cite{Isg:89,Rad:91}. It
is controversial whether or not (\ref{eq:DA-CZ}) is supported by
QCD sum rules. We do not want to enter that
discussion but consider (\ref{eq:DA-CZ}) as an example whose significance
is given by its frequent use.\\
{}From the $\pi^0\to\gamma\gamma$ constraint (\ref{eq:pimunupigamgam})
we find for the transverse size parameter $a$ the values $861\,{\rm MeV}$
for the asymptotic wave function and $673\,{\rm MeV}$ for the CZ wave
function. These values for $a$ correspond to the following characteristic
properties of the pion's valence Fock state: $0.21\,(0.32)\,{\rm fm}$ for
the radius, $367\,(350)\,{\rm MeV}$ for the root mean square
transverse momentum and $0.25 (0.32)$ for the probability when
the asymptotic (CZ) wave function is used. The small radius of the
valence Fock state has a lot of implications in hard
processes \cite{Fra:94}.\\
Numerical results for the transition form factor $F_{\pi\gamma}(Q^2)$
obtained from (\ref{eq:pigaff-ft}) are displayed in
Fig.~\ref{fig:pigaff}. We emphasize that there is no free parameter
to be adjusted once the wave function is
chosen. Obviously the results obtained from the asymptotic
wave function are in very good agreement with the CELLO data
\cite{Beh:91} as is also indicated by a $\chi^2$ value of 6.2 for
the five data points; there is not much room left for contributions from
higher order perturbative QCD and/or from higher Fock states. The results
obtained from the CZ wave function overshoot the data
significantly ($\chi^2=44.5$). Thus the comparison of both the results with
the data on the $\pi$-$\gamma$ transition form factor evidently ends
in favor of the asymptotic wave function. Of course, the limited quality
and quantity of the data allows mild modifications of the asymptotic wave
functions without worsening the agreement between theory and experiment
considerably. For example, if we follow Brodsky et al.\cite{Bro:83} and
multiply (\ref{eq:DA-AS}) by the exponential $\exp (-a^2 m_q^2/x_1x_2)$
where the parameter $m_q$ represents a constituent quark mass of, say,
$330\,{\rm MeV}$, we find similarly good results from this modified
asymptotic wave function (termed BHL subsequently) as from
(\ref{eq:DA-AS}) itself. On the other hand, strongly
end-point concentrated wave functions are clearly disfavored. Admixtures of
the second Gegenbauer polynomial enhancing the end-point regions, must
be small; already a value of 0.1 for the expansion coefficient $B_2$
deteriorates the agreement with the experimental data substantially
($\chi^2=10$).\\
The standard HSA \cite{Lep:80,Bro:81} predicts for the
$\pi$-$\gamma$ transition form factor
\begin{equation}
F_{\pi\gamma}(Q^2)= \sqrt{2}/3 \,f_\pi \langle x_1^{-1}\rangle\, Q^{-2}.
\end{equation}
The bracket term denotes the $x_1^{-1}$ moment of the distribution amplitude.
This moment receives the value 3 and 5 for the asymptotic and the CZ
distribution
amplitude respectively. Obviously, the standard HSA, while
exact at large $Q$, fails to describe the data in the few GeV region.
It does not provide the substantial $Q$-dependence the data for
$Q^2\,F_{\pi\gamma}$ exhibits. This is to be contrasted with the modified
HSA in which the QCD corrections, condensed in the Sudakov factor, and the
transverse degrees of freedom provide the
required $Q$-dependence. Asymptotically, i.~e., for
$\ln (Q/\mu_0) \to \infty$,
the Sudakov factor damps any contribution except those from
configurations with small quark-antiquark separation and, as the limiting
behavior, the QCD
prediction \cite{Bro:81,Wal:74} $F_{\pi\gamma}\to \sqrt{2}f_\pi\,Q^{-2}$
emerges.\\
The $Q$-dependence of the form factor can be parameterized as
\begin{equation}
F_{\pi\gamma}(Q^2)= A/(1+Q^2/s_0)
\label{eq:interpol}
\end{equation}
where for the constant $s_0$ a value of about $0.6-0.7\,{\rm GeV}^2$ is
required by the data. In vector meson dominance models $s_0$ is
to be identified with the square of the $\rho$-meson mass
($0.59\,{\rm GeV}^2$). Brodsky and Lepage \cite{Bro:81} propose
the formula (\ref{eq:interpol}) as an interpolation between the two limits,
$F_{\pi\gamma}(Q^2$=0$)=(2\sqrt{2}\,\pi^2f_\pi)^{-1}$ known from
current algebra and the above mentioned asymptotic behavior
$\sqrt{2}f_\pi\,Q^{-2}$, hence $s_0=4\pi^2{f_\pi}^2=0.67\,{\rm GeV}^2$. We
stress that this interpolation formula is not derived within the
standard HSA. In QCD sum rule analyses $s_0$ is related to the
pion's duality interval \cite{Rad:94}. Our results respect (\ref{eq:interpol})
approximately and $s_0$ represents the net effect of the suppressions
due to both the Sudakov factor and the $b_1$-dependence of the wave function
and of $\hat{T}_H$.\\
We close the discussion of the $\pi$-$\gamma$ transition form factor with
a reference to a paper by Szczepaniak and Williams \cite{Szc:94} in
which soft, nonperturbative corrections to the standard HSA are
estimated. These corrections bear resemblance to the intrinsic
transverse separation effects considered by us.

%%%%%%%%%%%%%%%%%%%%%%%%%%%%%%%%%%%%%%%%%%%%%%%%%%%%%%%%%%%%%%%%%%%%

\section{The $\eta$-$\gamma$ and $\eta^\prime$-$\gamma$
transition form factors}
\label{sec:etagaff}

We are now going to generalize (\ref{eq:pigaff-ft}) to the cases
of $\eta$-$\gamma$ and $\eta^\prime$-$\gamma$ transitions.
We start with the SU(3) basis states, $\eta_8$ and $\eta_1$, and employ the
usual mixing scheme
\begin{eqnarray}
\mid \eta \rangle &=&
 \cos \vartheta_P \mid \eta_8 \rangle -\sin \vartheta_P \mid \eta_1 \rangle
\nonumber\\
\mid \eta^\prime \rangle &=&
 \sin \vartheta_P \mid \eta_8 \rangle +\cos \vartheta_P \mid \eta_1 \rangle.
\label{eq:mixing}
\end{eqnarray}
Insertion of this scheme into the $\eta$-$\gamma$ and
$\eta^\prime$-$\gamma$
matrix elements of the electromagnetic current leads to relations between
the physical transition form factors and the $\eta_8$-$\gamma$ and the
$\eta_1$-$\gamma$ ones. The latter form factors can be calculated analogously
to the $\pi$-$\gamma$ case. For the $\eta_8$ and $\eta_1$ wave functions we
use the same ansatz as for the pion $(i=1,8)$
\begin{equation}
\hat\Psi_i (x_1,\vec b_1) = \frac{f_i}{2 \sqrt{6}}
\,\phi_i(x_1) \,\hat\Sigma_i(\sqrt{x_1x_2}\,b_1)
\label{eq:wvfct-ansatz-1-8}
\end{equation}
and assume that, except of the decay constants $f_i$, the two wave functions
are identical to the asymptotic pion wave function, (\ref{eq:Sigma-ft})
and (\ref{eq:DA-AS}). In the hard scattering amplitude (\ref{eq:pigaff-TH})
the charge factor of the pion is to be replaced by either
\begin{equation}
C_8=\left( {e_u}^2+{e_d}^2-2{e_s}^2 \right)/\sqrt{6}
\end{equation}
or
\begin{equation}
C_1=\left( {e_u}^2+{e_d}^2+{e_s}^2 \right)/\sqrt{3}.
\label{eq:singlet-charge}
\end{equation}
Furthermore we correct for the rather large masses of
the $\eta_i$-mesons ($m_8=566\,{\rm MeV}$, $m_1=947\,{\rm MeV}$,
see \cite{Leu:85}). Hence, in the transverse separation space, the hard
scattering amplitude reads
\begin{equation}
\hat T_H(x_1,\vec b_1,Q) =
\frac{2\sqrt{6}\,C_i}{\pi}
K_0(\sqrt{x_2Q^2+x_1x_2{m_i}^2}\;b_1).
\label{eq:etagaff-TH-ft}
\end{equation}
Inserting (\ref{eq:wvfct-ansatz-1-8}) and (\ref{eq:etagaff-TH-ft}) into
(\ref{eq:pigaff-ft}) we can compute the $F_{\eta_i\gamma}$ form factors
and, then, using the mixing scheme (\ref{eq:mixing}), the
$\eta$-$\gamma$ and $\eta^\prime$-$\gamma$ transition form factors. However,
we also need for this calculation the values of the decay constants and
the mixing angle. Since these par\-ameters are not known with sufficient
accuracy we will change our attitude and, encouraged by the success of the
modified HSA in the $\pi$-$\gamma$ case, try to determine them. Admittedly,
additional information is required for this task since the $\eta$-$\gamma$
and $\eta^\prime$-$\gamma$ transition form factors do not suffice to fix the
three parameters; for any value of the mixing angle a reasonable fit to the
data is obtained. The necessary extra information is provided by
the two-photon
decays of the $\eta$ and $\eta^\prime$. Adapting the PCAC result for the
$\pi^0\to\gamma\gamma$ decay to the $\eta$ and $\eta^\prime$ case with proper
mixing at the amplitude level, one finds for the decay widths
\begin{eqnarray}
\Gamma(\eta\to\gamma\gamma )&=&
\frac{9\alpha^2}{16\pi^3}{m_\eta}^3
\left[\frac{C_8}{f_8}\cos\vartheta_P -\frac{C_1}{f_1}\sin\vartheta_P\right]^2
\nonumber \\
\Gamma(\eta^\prime\to\gamma\gamma )&=&
\frac{9\alpha^2}{16\pi^3}{m_{\eta^\prime}}^3
\left[\frac{C_8}{f_8}\sin\vartheta_P +\frac{C_1}{f_1}\cos\vartheta_P\right]^2
\label{eq:decaywidths}
\end{eqnarray}
The experimental values for the decay widths are \cite{PDG:94}:
\begin{equation}
\Gamma (\eta\to\gamma\gamma )=0.510\pm 0.026\,{\rm keV}
\qquad\qquad
\Gamma (\eta^\prime\to\gamma\gamma )=4.26\pm 0.19\,{\rm keV}
\end{equation}
For obvious reasons we have only quoted the PDG average
of the two-photon measurements of the $\eta\to\gamma\gamma$ width. The value
of $0.324\pm 0.046\,{\rm keV}$ obtained from the Primakoff production
measurement \cite{Bro:74} will not be used in our analysis.\\
We evaluate the three parameters, $f_1$, $f_8$ and $\vartheta_P$, through a
combined least square fit to the data on the form factors and the decay
widths. The parameters acquire the following values:
\begin{equation}
f_1\,=145\pm 3\,{\rm MeV},\qquad f_8\,=136\pm 10\,{\rm MeV},\qquad
\vartheta_P\,=-18^\circ \pm 2^\circ
\end{equation}
and the $\chi^2$ value is 14.8 for the 18 data points.
The values of the parameters and their errors are only correct provided the
$\eta$ and $\eta^\prime$ wave functions are at least approximately given by
(\ref{eq:wvfct-ansatz-1-8}), (\ref{eq:Sigma-ft}) and (\ref{eq:DA-AS}).
Since $f_1$ and $f_8$ have rather similar values nonet symmetry of the
wave functions holds approximatively. The decay constants of the physical
mesons are:
\begin{equation}
f_\eta\,=175\pm 10\,{\rm MeV};
\qquad\qquad f_{\eta^\prime}\,=95\pm 6\,{\rm MeV}.
\end{equation}
The quality of the fit can be judged from Fig.~\ref{fig:etagaff}
where fit and data
\cite{Ber:84,Aih:90,Beh:91} for the transition form factors are shown. As
expected from the very good value of the $\chi^2$ the agreement between theory
and experiment is excellent. The computed values for the decay widths are
$\Gamma\,(\eta\to\gamma\gamma)\,=0.5\,{\rm keV}$ and
$\Gamma\,(\eta^\prime\to\gamma\gamma)\,=4.17\,{\rm keV}$.\\
Our value for the mixing angle is compatibel with other results. For instance,
a value of $ -20^\circ \pm 4^\circ $ is obtained from chiral perturbation
theory \cite{Leu:85}. An analysis of two-meson final states produced in
proton-antiproton annihilations seems to provide a similar value for
the mixing angle ($-17.3^\circ \pm 1.8^\circ $) \cite{Ams:92}. Gilman
and Kauffman
\cite{Gil:87} found from a phenomenological analysis of various decay processes
and of $\pi^{-}p$ scattering that a value of about $-20^\circ$ is favored.
{}From a similar analysis but under inclusion of constituent quark mass effects
Bramon and Scadron \cite{Bra:90} obtained
$\vartheta_P=\,-14^\circ\pm 2^\circ$.\\
{}From chiral perturbation theory Gasser and Leutwyler \cite{Leu:85} predicted
a value of $170\pm 7\,{\rm MeV}$  for the $\eta_8$ decay constant whereas
Donoghue et al.~\cite{Don:85} found $163\,{\rm MeV}$. While both the values
are compatible within the uncertainties of the chiral perturbation theory,
they are larger than our result. In order to see
whether or not such large value is definitively excluded in our approach we
repeat the combined fit, keeping $f_8$ at the value of $163\,{\rm MeV}$. The
resulting fit
is not as good as the precedent fit but still of acceptable quality, the
value of $\chi^2$ is 20.7. It provides:
$f_1\,=143\pm 3\,{\rm MeV}$, $\vartheta_P\,=-21^\circ \pm 1^\circ$
and hence
$f_\eta\,=203\pm 2\,{\rm MeV}$, $f_{\eta^\prime}\,=76\pm 5\,{\rm MeV}$.
The results for the form factors are almost as good as before. The resulting
decay widths are $\Gamma\,(\eta\to\gamma\gamma)\,=0.47\,{\rm keV}$ and
$\Gamma\,(\eta^\prime\to\gamma\gamma)\,=4.20\,{\rm keV}$, i.~e., imposing
larger values upon $f_8$ forces the two-photon decay width of the $\eta$
towards smaller values closer to the result of the Primakoff experiment
\cite{Bro:74}. In view of the experimental uncertainties in the
$\eta\to\gamma\gamma$ and of the moderate difference in the $\chi^2$ we
cannot exclude the possibility of a $f_8$ as large as $163\,{\rm MeV}$
although a value around $140\,{\rm MeV}$ is favored from our analysis.
In contrast to $f_8$ the other two parameters, $f_1$ and $\vartheta_P$,
are tightly constrained. We stress that the quoted errors
are only those obtained in the statistical analysis. By no means they
reflect the full uncertainties of the parameters which are rather represented
by the differences between the two sets of parameters. In
\cite{Aih:90,Beh:91} the $\eta$ and $\eta^\prime$ decay constants
have been determined from fitting pole formul{\ae} analogous to
(\ref{eq:interpol}) to the transition form factor data. The values obtained
for the decay constants differ from our ones considerably.
Mixing is not taken into account in these analyses.\\
It is often speculated upon a gluon admixture to the $\eta_1$
\begin{equation}
\eta_1=\left[ u \bar u+d \bar d+s\bar s+\epsilon G
\right]/\sqrt{3+\epsilon^2}.
\end{equation}
Allowing for that component which only changes the
singlet charge factor (\ref{eq:singlet-charge}) through the normalization,
and repeating the fits to the data, we do not find any evidence for a
sizeable gluon admixture to the $\eta_1$ ($\epsilon = 0 \pm 0.2$).\\
Finally we would like to comment on another mixing scheme frequently
discussed in the literature (see, for instance, \cite{Gil:87,Bra:90,Fri:77}),
in which one uses quark states, $u\bar u+d\bar d$ and $s\bar s$, as
the basis. Along the same lines as discussed above one may calculate
the transition form factors using this mixing scheme. However, since
we do not know well enough the masses and the wave functions of the
quark basis states we refrain from carrying out that analysis.

%%%%%%%%%%%%%%%%%%%%%%%%%%%%%%%%%%%%%%%%%%%%%%%%%%%%%%%%%%%%%%%%%%%%%

\section{do we know the pion's wave function?}
\label{sec:piff}

In this section we are going to put together the available information upon
the pion's wave function. For this purpose we call to mind
previous analyses of the elastic form factor of the pion and of its
valence quark distribution function and discuss the results of these analyses
in the light of our observations made on the $\pi$-$\gamma$ transition form
factor.\\
Within the modified HSA the pion form factor is to be calculated
from the relation \cite{LiS:92,Li:92,Jak:93}
\begin{eqnarray}
{F_\pi}^{pert}(Q^2)&=&
\int_0^1\! \frac{dx_1 \, dy_1}{(4 \pi)^2}
\int_{-\infty}^\infty \!d^{\;\!2}b_1 \; d^{\;\!2}b_2
\,\hat{\Psi}_0^\ast (y_1,\vec b_2)\,\hat{T}_H (x_1,y_1,Q,\vec b_1,\vec b_2,t)
\,\hat{\Psi}_0 (x_1,-\vec b_1) \nonumber \\
& &\qquad\qquad\qquad
\times\,\exp\left[-S(x_1,Q,b_1,t)-S(y_1,Q,b_2,t)\right]
\label{eq:piff-ft}
\end{eqnarray}
where the hard scattering amplitude $\hat T_H$ is given by the
product of two zeroth order Bessel functions arising from the
gluon and the quark propagators in the elementary Feynman diagrams
\begin{equation}
\hat T_H(x_1,y_1,Q,\vec b_1,\vec b_2)=
\frac{4\alpha_s(t)C_F}{\pi}\;
K_0(\sqrt{x_1y_1}Qb_2)\;
K_0(\sqrt{x_1}Q\,|\vec b_1+\vec b_2|)
\label{eq:piff-TH-ft}
\end{equation}
$C_F(=4/3)$ is the color factor and the Sudakov exponents $S$
are defined in (\ref{eq:pigaff-sudakov}). $t$, the largest
mass scale appearing in the hard scattering amplitude, is now given by
$\max(\sqrt{x_1y_1}Q,1/b_1,1/b_2)$. Equation (\ref{eq:piff-ft})
implies two angle integrations, say the integration over the direction
of $\vec b_2$ which simply provides a factor of $2\pi$, and the
integration over the relative angle $\phi$ between $\vec b_1$ and
$\vec b_2$. The latter integration can be carried out analytically
by means of Graf's theorem
\begin{equation}
\frac{1}{2\pi} \int d\varphi \;
K_0(c\,|\vec b_1+\vec b_2|)=
 \theta (b_1-b_2)K_0(c\,b_1)I_0(c\,b_2)
+\theta (b_2-b_1)K_0(c\,b_2)I_0(c\,b_1),
\label{eq:grafstheorem}
\end{equation}
where $\theta$ is the usual step function and $I_0(x)=J_0(ix)$;
$J_0$ is the Bessel function of order zero. Thus a four dimensional
integration remains to be carried out numerically.\\
The perturbative contributions to the elastic form factor obtained from both
the wave functions, the asymptotic and the CZ one, are compared to
the data \cite{Beb:76} in Fig.~\ref{fig:piff}. Both the predictions, while
theoretically self-consistent for momentum transfers larger than about
$2\,{\rm GeV}$, fall short of the data. One may wonder about the reason
for the failure of the modified HSA in the case of the elastic form factor
while it works so well for the $\pi$-$\gamma $ transition form factor. At this
point we remind the reader that the elastic form factor also gets a
contribution from the overlap of the initial and final state soft wave
functions $\hat\Psi_0$ \cite{Dre:70}. A contribution
of this type does not exist for the
transition form factor. Formally the perturbative contribution to the
elastic form factor represents the overlap of the large transverse
momentum ($k_\perp$) tails of the wave functions while the overlap
of the soft parts of the wave functions is assumed to be negligible
small at large $Q$. Since the soft wave functions are explicitly used
in applications of the modified HSA it is natural, even obligatory, to
examine the validity of that presumption. If the overlap of the soft wave
functions turns out to be of substantial magnitude it must be taken
into account in the calculation of the elastic form factor for consistency!
According to \cite{Dre:70} the overlap of the soft transverse configuration
space wave functions reads
\begin{equation}
{F_\pi}^{soft}(Q^2)=
\frac{1}{4\pi}\int dx_1
\int d^{\;\!2}b_1
\exp\left[-(1-x_1)\vec b_1\cdot \vec q_\perp\right]\;
|\hat\Psi_0(x_1,\vec b_1)|^2,
\label{eq:piff-soft-ft}
\end{equation}
where $Q^2={\vec {q}_\perp}^{\;\!2}$. As the inspection of
(\ref{eq:piff-soft-ft}) reveals only the end-point region
$1\geq x_1 \gsim 1- 1/(Q\langle b_1^2\rangle ^{1/2})$ contributes
at large $Q$,
i.~e., only configurations where the photon interacts with a parton
carrying almost the entire momentum of the pion. Higher Fock states
provide similar overlap terms. Obviously with an increasing number of partons
sharing the pion's momentum it becomes less likely that one parton carries
the full momentum of the pion. Therefore higher Fock state overlap terms
are strongly suppressed at large $Q$. Consider for example the $q\bar{q} g$
Fock state and assume that the corresponding wave function is a suitable
3-particle generalization of (\ref{eq:wvfct-ansatz}), (\ref{eq:Sigma-ft}) and
(\ref{eq:DA-AS}) where the asymptotic form of the $q\bar{q} g$ distribution
amplitude is $\sim x_1x_2x_3^2$ ($x_3$ refers to the gluon) \cite{Che:84}.
It is easy to convince oneself that the $q\bar{q} g$ overlap term behaves as
$1/Q^{12}$ at large $Q$; above $Q\simeq2\,{\rm GeV}$ it can be neglected to
any degree of accuracy.\\
Soft contributions calculated from (\ref{eq:piff-soft-ft}) are shown in
Fig.~\ref{fig:piff}. The results obtained from the asymptotic
wave function obviously have the right magnitude to fill the gap
between the corresponding perturbative results
and the data.\footnote{We refrain from showing the sums of soft and
perturbative contributions because some double-counting might be implied.}
The broad flat maximum of the overlap contribution simulates a $Q$-dependence
of the form factor which, in the few GeV region, resembles the $Q$-dependence
predicted by the dimensional counting rules. For the CZ wave function
the overlap contribution
exceeds the data significantly; the maximum value of $Q^2 F_\pi$ amounts to
$2.1 \,{\rm GeV}^2$ and is located at $Q=7.4\,{\rm GeV}$. We consider this
result as a serious failure of the CZ wave function.\\
At large $Q$ the overlap contributions from our wave functions are
suppressed by $1/Q^2$ as compared to the perturbative contribution.
At which value of momentum transfer the transition from the dominance
of soft to hard contributions actually happens depends on the end-point
behavior of the wave function sensitively. Already a moderate modification
of the wave function may shift that value considerably. For the asymptotic
wave function the transition takes place at the rather large value
of about $10\,{\rm GeV}$ while for the BHL wave function the additional
exponential effectuates a sharper decrease of the soft contribution
and the hard contribution takes the control at about $5\,{\rm GeV}$
(see Fig.~\ref{fig:piff}). However, below $3\,{\rm GeV}$ the
contributions from both the
wave functions do not differ much. They also provide similar results for the
$\pi$-$\gamma$ form factor as we already mentioned.\\
Large overlap contributions have also been observed by other authors
\cite{Jak:93,Isg:89,Kis:93,Chu:88}. Thus the small size of the
perturbative contribution finds a comforting
explanation a fact which has already been pointed out by Isgur and
Llewellyn-Smith \cite{Isg:89}. Were that contribution (including higher
order perturbative corrections) much larger the existing large overlap
contributions would be inexplicable. This conclusion seems to be compatibel
with calculations of the one-loop corrections to the perturbative contribution
\cite{Fie:81,Dit:81}.\\
The structure function of the pion or rather its parton
distribution functions $q_\nu(x)$ offer another possibility to test
our wave functions against data. As
has been discussed in \cite{Bro:83} and \cite{Hua:94} the
parton distribution functions are determined  by the Fock state
wave functions. Each Fock state contributes through the modulus
squared of its wave function integrated over transverse momenta
up to $Q$ and over all fractions $x_i$ except
those pertaining to partons of the type $\nu$. Obviously the valence Fock
state wave function only feeds the valence quark
distribution function $q_d^V=q_{\bar u}^V$. Since each
Fock state contributes to the distribution functions
positively, the inequality
\begin{equation}
q_d^V(x,Q)\geq
\int dx_1 \int^Q \frac{d^{\;\!2}k_\perp}{16\pi^3}\;
|\Psi(x_i,\vec k_\perp)|^2 \;
\delta(x_1-x)
\label{eq:quarkdistribution}
\end{equation}
holds. In order to calculate the $Q$ dependence of the valence quark
distribution function we need to know the full pion wave function, its
perturbative tail is responsible for the evolution behavior. Yet the soft
wave function (\ref{eq:wvfct-ansatz}) provides the bulk of the valence Fock
state contribution to the distribution function and, consequently, in the
few GeV region this contribution should respect the inequality
(\ref{eq:quarkdistribution}). Hence
\begin{equation}
q_d^V(x)\geq
\frac{\pi^2}{3}\, f_\pi^2 \, a^2 \,
\frac{\phi^2(x)}{x(1-x)}.
\end{equation}
Since, on the average, higher Fock state partons possess smaller values
of $x$ than partons from the valence Fock state
the inequality should become an equality for $x\to 1$. For other values of
$x$ many other Fock states contribute to the distribution function in
general, a fact which has to be contrasted with exclusive reactions to
which only the valence Fock state contributes at large $Q$.\\
In Fig.~\ref{fig:distribution} the results for the valence quark structure
function obtained from the asymptotic and from the CZ wave functions
are shown and compared to the parameterization
\begin{equation}
x\, q_d^V(x)=A\, x^{\delta_1}(1-x)^{\delta_2}.
\label{eq:val}
\end{equation}
The constant $A$ is determined by the requirement
$\int_0^1 q_d^V(x)\;dx=1$. Sutton et al. \cite{Sut:92} analysed the recent
data on the Drell-Yan process $\pi^-p\to\mu^+\mu^-X$ and determined the
values of the other two parameters; they quote $\delta_1=0.64\pm 0.03$
and (the averaged value) $\delta_2=1.11\pm0.06$ at $Q= 2\,{\rm GeV}$. In
Fig.~\ref{fig:distribution} the parameterization of \cite{Sut:92} is
displayed as a band indicating the uncertainties of it. The width
of the band does not take into account the uncertainties induced by
theoretical assumptions underlying that analysis (e.~g., a $K$
factor). In particular near $x=1$ the width of the
band may be underestimated since the parameterization of the proton
structure function, used as input in that analysis, is an extrapolation
for $x\gsim 0.75$, i.~e., it is not supported by data.\\
The comparison of the parameterization (\ref{eq:val}) with the predictions
obtained from our wave functions clearly reveals that the asymptotic wave
function respects the inequality (\ref{eq:quarkdistribution}). The little
excess around $x=0.9$ is perhaps a consequence of the above mentioned
extra uncertainties in the parameterization (\ref{eq:val}) and/or of
small deviations of the wave function from the asymptotic form. On the
other hand, the CZ wave function exceeds the result of \cite{Sut:92}
dramatically at large $x$. The above-mentioned uncertainties
can not account for that. Thus, the CZ wave function fails again and
is, therefore, to be rejected. This conclusion has already been drawn by
Huang et al. \cite{Hua:94}. However, our analysis of the $\pi$-$\gamma$
transition form factor, a quantity which has not been considered in
\cite{Hua:94}, strengthens the evidence for the asymptotic wave function
and against the CZ one.\\
Frequently the pion form factor is calculated via the overlap formula
(\ref{eq:piff-soft-ft}) using a wave function normalized to
unity (e.~g., \cite{Kis:93,Chu:88}). It is customarily asserted that
such a wave
function describes the binding of constituent quarks in a pion.
While such a wave function may provide an overlap
contribution to the pion form factor in fair agreement with the data
at all $Q^2$ and may also respect the $\pi\to\mu\nu$ constraint it is
hard to see how a conflict with the data for the other three reactions
we are considering, can be avoided. In order to illustrate the
difficulties arising from a wave function normalized to unity, let us look
to the asymptotic wave function again. Duplication of the parameter
$a$ normalizes the wave function to unity and then the following results are
obtained: the contribution to the valence quark distribution function is
four times bigger than that one shown in Fig.~\ref{fig:distribution}, the
$\pi$-$\gamma$ form factor becomes too large ($\chi^2=49$)
and the $\pi\to\gamma\gamma$ constraint is badly violated. The pion form
factor on the other hand is well described. We finally note that
the BHL wave function (normalized to unity) leads to results for the elastic
form factor which are very similar to those presented in \cite{Kis:93}.
Yet the deficiencies in the other reactions remain.

%%%%%%%%%%%%%%%%%%%%%%%%%%%%%%%%%%%%%%%%%%%%%%%%%%%%%%%%%%%%%%%%%%%%%%%%%%%

\section{conclusions}
\label{sec:concl}

We summarize our findings about reactions involving pions as the only
hadrons: The asymptotic wave function as the only phenomenological
input leads to a successful description of four exclusive processes
and is compatible with the band on the pion's valence quark
distribution function. The four processes are, on the one side, the
leptonic decay of the charged pion and the two-photon decay of the
uncharged pion fixing the parameters of the wave function and, on the
other side, the $\pi$-$\gamma$ transition form factor as well as the pion's
elastic form factor which are calculated in the few GeV region within
the modified HSA. This success nicely demonstrates the universality of
the pion wave function, i.~e., its process independence. Of course, the
present quality of the form factor data does not pin down the form of
the wave function precisely. Wave functions close to the asymptotic form
are compatible with the data likewise. The BHL form is an example of such a
wave function. On the other hand, the CZ wave function as well as
other strongly end-point concentrated wave functions are clearly
in conflict with the data and should therefore be discarded. The use of such
wave functions in the analyses of other exclusive
reactions, e.~g., $\gamma\gamma\to\pi\pi$ or $B\to\pi\pi$, is
unjustified and likely leads
to overestimates of the perturbative contributions.\\
We also analyzed the $\eta$-$\gamma$ and $\eta^\prime$-$\gamma$ transition
form factors within the modified HSA. Assuming for the SU(3) basis states
the same asymptotic wave functions as for the pion, we determined the decay
constants and the mixing angle from a combined fit to the data on form
factors and decay widths and found: $f_\eta\,=175\,{\rm MeV}$,
$f_{\eta^\prime}\,=95\,{\rm MeV}$ and
$\vartheta_P\,=-18^\circ$. These values correspond to
$f_8=145\,{\rm MeV}$ which is smaller than the chiral perturbation theory
prediction. Keeping $f_8$ at the predicted value of $163\,{\rm MeV}$, we found:
$f_\eta\,=203\,{\rm MeV}$, $f_{\eta^\prime}\,=76\,{\rm MeV}$ and
$\vartheta_P\,=-21^\circ$. Owing to the experimental
uncertainties in the $\eta\to\gamma\gamma$ and of the moderate difference
in the $\chi^2$ we cannot exclude the second set of parameters
although the first set is favored from our analysis.\\
The parameters are determined under the assumption that the asymptotic wave
function is close to reality. Since our values of the parameters are
fairly compatible with those found in other analyses, we are tempted to
conclude that the $\eta_8$ and $\eta_1$ wave functions we use, are indeed
approximately correct.

%%%%%%%%%%%%%%%%%%%%%%%%%%%%%%%%%%%%%%%%%%%%%%%%%%%%%%%%%%%%%%%%%%%%%%%%%%%
\acknowledgments
We thank M.~Feindt, H.~Fritzsch and H.~Genz for helpful discussions. This
work was partially supported by the Bundesministerium f\"ur Forschung
und Technologie and by the Deutsche Forschungsgemeinschaft.\\

% \begin{figure}
% \caption{}
% \label{}
% \end{figure}

\begin{figure}
\caption[dummy1]{The basic diagrams for the meson-photon
transition form factor.}
\label{fig:diags}
\end{figure}

\begin{figure}
\caption[dummy2]{The $\pi$-$\gamma$ transition form factor vs.~$Q^2$.
The solid (dashed) line represents the prediction obtained with the
modified HSA using the asymptotic (CZ) wave function. The dotted line
represents the results of the standard HSA (for the asymptotic wave function).
Data are taken from \cite{Beh:91}.}
\label{fig:pigaff}
\end{figure}

\begin{figure}
\caption[dummy3]{The $\eta$-$\gamma$ and $\eta^\prime$-$\gamma$ transition
form factors vs.~$Q^2$. The solid lines represent the predictions obtained
from the modified HSA using the asymptotic wave function. Data are taken
from  PLUTO \cite{Ber:84} ($\,$\rule{2.3mm}{2.3mm}$\,$),
TPC/$2\gamma$ \cite{Aih:90}
({\large $\bullet$}) and CELLO \cite{Beh:91} ({\large $\circ$}).}
\label{fig:etagaff}
\end{figure}

\begin{figure}
\caption[dummy4]{The elastic form factor of the pion vs. $Q^2$. The
dash-dotted (dash-dot-dotted) line represents the perturbative
contributions obtained from (\ref{eq:piff-ft}) using the asymptotic
(CZ) wave function. The solid (dashed) line represents the overlap contribution
obtained from the soft part of the asymptotic (CZ) wave function.
For comparison the overlap contribution obtained from the BHL
wave function is also shown (dotted line). Data are taken from \cite{Beb:76}.}
\label{fig:piff}
\end{figure}

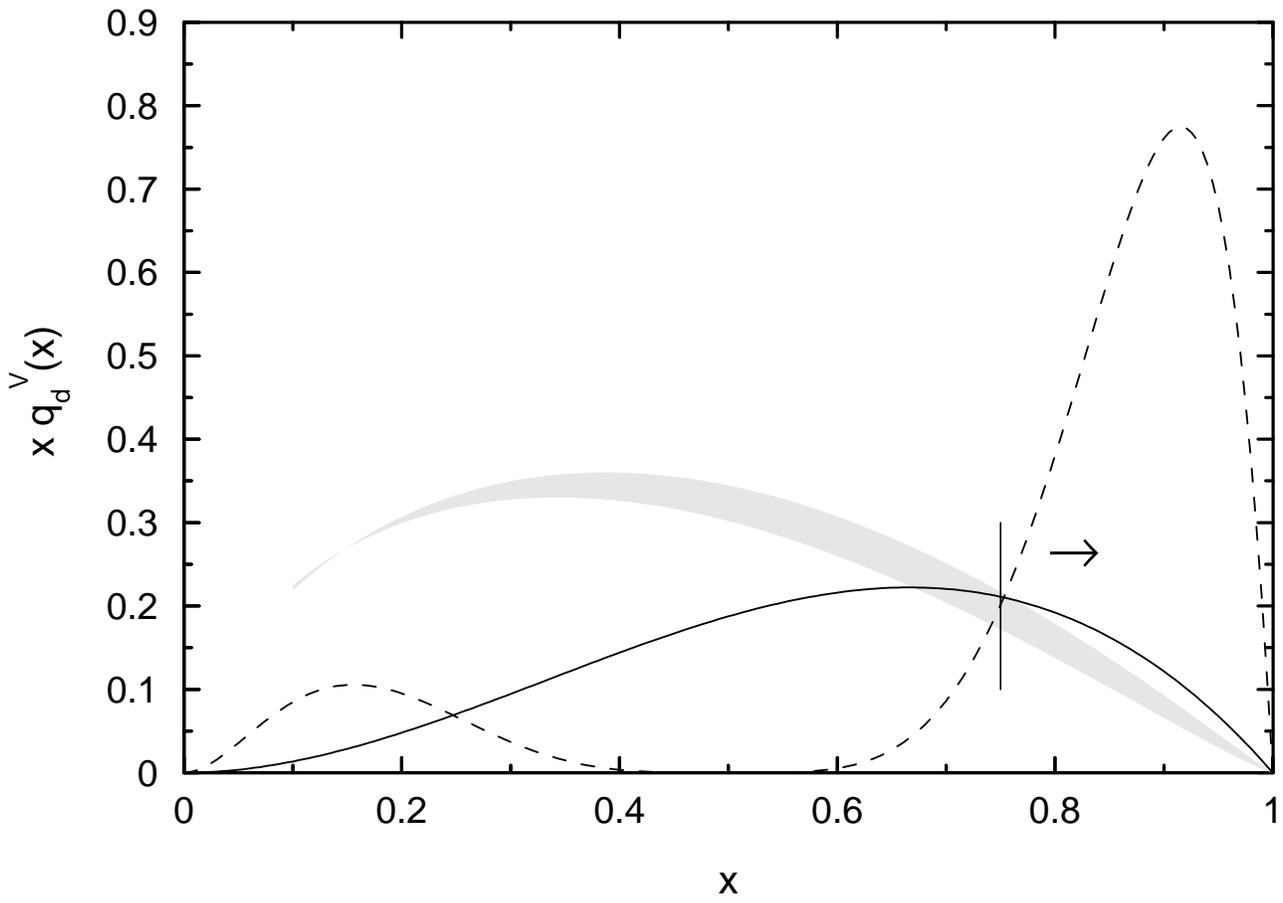
\begin{figure}
\caption[dummy5]{The valence quark distribution function $xq_d^V(x)$ at
$Q^2\simeq5\,{\rm GeV}^2$. The shadowed band represents the parameterization
by \cite{Sut:92}. For other symbols refer to Fig.~\ref{fig:pigaff}. The
bar and the arrow indicate the range $x$ in which an extrapolation for
the proton structure function is used.}
\label{fig:distribution}
\end{figure}

\newpage

\begin{references}
%
\bibitem{Bot:89}
J.~Botts and G.~Sterman, Nucl.~Phys.~B325~(1989)~62.
%
\bibitem{LiS:92}
H.~N.~Li and G.~Sterman, Nucl.~Phys.~B381~(1992)~129.
%
\bibitem{Li:92}
H.~N.~Li, Phys.~Rev.~D48~(1993)~4243.
%
\bibitem{Lep:80}
G.~P.~Lepage and S.~J.~Brodsky, Phys.~Rev.~D22~(1980)~2157.
%
\bibitem{Jak:93}
R.~Jakob and P.~Kroll, Phys.~Lett.~B315~(1993)~463; B319~(1993)~545(E).
%
\bibitem{Bol:94}
J.~Bolz, R.~Jakob, P.~Kroll, M.~Bergmann and N.G.~Stefanis,
preprints WU-B-94-06 and WU-B-94-16, Wuppertal (1994).
%
\bibitem{Dre:70}
S.~D.~Drell and T.~M.~Yan, Phys.~Rev.~Lett.~24~(1970)~181;\\
G.~West, Phys.~Rev.~Lett.~24~(1970)~1206.
%
\bibitem{Ber:84}
PLUTO coll., Ch. Berger et al., Phys.~Lett.~B142~(1984)~125.
%
\bibitem{Aih:90}
TPC/2$\gamma$ coll., H.~Aihara et al., Phys.~Rev.~Lett.~64~(1990)~172.
%
\bibitem{Beh:91}
CELLO coll., H.-J.~Behrend et al., Z.~Phys.~C49~(1991)~401.
%
\bibitem{Mue:94}
C.R.~M\"unz, J.~Resag, B.C.~Metsch and H.R.~Petry,
preprint Bonn (1994).
%
\bibitem{Bro:83}
S.~J.~Brodsky, T.~Huang and G.~P.~Lepage,
Banff Summer Institute, Particles and Fields~2, p.~143,
A.Z.~Capri and A.N.~Kamal (eds.), 1983.
%
\bibitem{Zhi:93}
A.~R.~Zhitnitsky, Phys.~Lett.~B329~(1994)~493 and
preprint SMU-HEP-94-19 (1994).
%
\bibitem{Che:82}
V.~L.~Chernyak and A.~R.~Zhitnitsky, Nucl.~Phys.~B201~(1982)~492.
%
\bibitem{Isg:89}
N.~Isgur and C.~H.~Llewellyn~Smith, Nucl.~Phys.~B317~(1989)~526.
%
\bibitem{Rad:91}
A.~V.~Radyushkin, Nucl.~Phys.~A532~(1991)~141c and references therein.
%
\bibitem{Fra:94}
L.~L.~Frankfurt, G.~A.~Miller and M.~Strikman, preprint DOE/ER/40427-06-N94
          (1994).
%
\bibitem{Bro:81}
S.~J.~Brodsky and G.~P.~Lepage, Phys.~Rev.~D24~(1981)~1808.
%
\bibitem{Wal:74}
T.~F.~Walsh and P.~Zerwas, Nucl.~Phys.~B41~(1972)~551.
%
\bibitem{Rad:94}
A.~V.~Radyushkin, preprint CEBAF-TH-94-15, Newport News (1994).
%
\bibitem{Szc:94}
A.~Szczepaniak and A.~G.~Williams, preprint ADP-93-216/T134
%
\bibitem{Leu:85}
J.~Gasser and H.~Leutwyler, Nucl.~Phys.~B250~(1985)~465.
%
\bibitem{PDG:94}
Particle Data Group, Phys.~Rev.~D50~(1994)~1173.
%
\bibitem{Bro:74}
A.~Browman et al., Phys.~Rev.~Lett.~33~(1974)~1400.
%
\bibitem{Ams:92}
Cristal Barrel coll., C.~Amsler et al., Phys.~Lett.~B294~(1992)~451.
%
\bibitem{Gil:87}
F.~J.~Gilman and R.~Kauffman, Phys.~Rev.~D36~(1987)~2761.
%
\bibitem{Bra:90}
A.~Bramon and M.~D.~Scadron, Phys.~Lett.~B234~(1990)~346.
%
\bibitem{Don:85}
J.~F.~Donoghue, B.~R.~Holstein and Y.-C.~R.~Lin,
Phys.~Rev.~Lett.~55~(1985)~2766.
%
\bibitem{Fri:77}
H.~Fritzsch and J.~D.~Jackson, Phys.~Lett.~B66~(1977)~365.
%
\bibitem{Beb:76}
C.~J.~Bebek et al., Phys.~Rev.~D13~(1976)~25 and D17~(1978)~1693.
%
\bibitem{Che:84}
V.~L.~Chernyak and A.~R.~Zhitnitsky, Phys.~Rep.~C112~(1984)~173.
%
\bibitem{Kis:93}
L.~S.~Kisslinger and S.~W.~Wang, Nucl.~Phys.~B399~(1993)~63.
%
\bibitem{Chu:88}
P.~L.~Chung, F.~Coester and W.~N.~Polyzou, Phys.~Lett.~B205~(1988)~545.
%
\bibitem{Fie:81}
R.~D.~Field et al., Nucl.~Phys.~B186~(1981)~429.
%
\bibitem{Dit:81}
F.~M.~Dittes and A.~V.~Radyuskin, Sov.~J.~Nucl.~Phys.~34~(1981)~293.
%
\bibitem{Hua:94}
T.~Huang, B.-Q.~Ma and Q-X.~Sheng, Phys.~Rev.~D49~(1994)~1490.
%
\bibitem{Sut:92}
P.~J.~Sutton, A.~D.~Martin, R.G.~Roberts and W.~J.~Stirling,
Phys.~Rev.~D45~(1992)~2349.
%

\end{references}
\end{document}